\def\tr{\hbox{tr}}
\def\ln{\ell{n}}
\documentstyle[a4,12pt]{article}
\begin{document}
\begin{titlepage} \vspace{0.2in} \begin{flushright}
MITH-97/10 \\ \end{flushright} \vspace*{1.5cm}
\begin{center} {\LARGE \bf  Further discussions on a possible lattice chiral 
gauge theory\\} \vspace*{0.8cm}
{\bf She-Sheng Xue$^{(a)}$}\\ \vspace*{1cm}
INFN - Section of Milan, Via Celoria 16, Milan, Italy\\ \vspace*{1.8cm}
{\bf   Abstract  \\ } \end{center} \indent
 
In a possible $SU_L(2)$ lattice chiral gauge theory with a large multifermion
coupling, we try to further clarify the threshold phenomenon: the possibility
that the right-handed three-fermion state turns into the virtual states of its
constituents (free chiral fermions) in the low-energy limit. Provided this
phenomenon occurs, we discuss the chiral gauge coupling, Ward identities and the
gauge anomaly within the gauge-invariant prescription of the perturbative
chiral gauge theory. 

\vfill \begin{flushleft}  June, 1996 \\
PACS 11.15Ha, 11.30.Rd, 11.30.Qc  \vspace*{3cm} \\
\noindent{\rule[-.3cm]{5cm}{.02cm}} \\
\vspace*{0.2cm} \hspace*{0.5cm} ${}^{a)}$ 
E-mail address: xue@milano.infn.it\end{flushleft} \end{titlepage}
 
\noindent
{\bf 1.}\hskip0.3cm
Since the ``no-go'' theorem \cite{nn81} of Nielsen and Ninomiya was
demonstrated in 1981, the problem of chiral fermion ``doubling'' and
``vector-like'' feature on a lattice still exists if one insists on
preserving chiral gauge symmetries. Actually, the essential spirit of the
``no-go'' theorem of Nielsen and Ninomiya is that, under certain prerequisites,
the paradox concerning chiral gauge symmetries, vector-like doubling and
anomalies are unavoidable. 

Eichten and Preskill\cite{ep} proposed a model of multifermion couplings ten
years ago. The crucial points of multifermion coupling can be briefly described
as follows. Multifermion couplings are introduced such that, in the 
strong-coupling phase, Weyl states composing three elementary Weyl fermions
(three-fermion states) are bound. Then, these three-fermion states pair up with
elementary Weyl fermions to become Dirac fermions. Such Dirac fermions may be
massive without violating chiral gauge symmetries, due to the appropriate
quantum numbers and chirality carried by these three-fermion states. The
binding thresholds of such three-fermion states depend on elementary Weyl modes
residing in different regions of the Brillouin zone. If one assumes that the
spontaneous symmetry breaking of the Nambu-Jona Lasinio (NJL) type \cite{njl}
does not occur and such binding thresholds separate the weak-coupling symmetric
phase from the strong-coupling symmetric phase, there are two possibilities to
realize the continuum limit of chiral fermions in coupling space. One is to
cross the binding threshold of the three-fermion state of chiral
fermions; another is of a wedge between two thresholds, where the three-fermion
state of chiral fermions has not been formed, provided all doublers sitting in
various edges of the Brillouin zone have been bound to be massive Dirac
fermions and decouple. 

The model\cite{ep} was studied in ref.~\cite{gpr}, where it was pointed out
that such a model of multifermion couplings fails to give chiral fermions in the
continuum limit. The reasons are that an NJL spontaneous symmetry-breaking
phase separating the strong-coupling symmetric phase from the weak-coupling
symmetric phase, the right-handed Weyl states do not completely disassociate
from the left-handed chiral fermions and the phase structure of such a model of
multifermion couplings is similar to that of the Smit-Swift (Wilson-Yukawa)
model\cite{ss}, which has been very carefully studied and shown to 
fail\cite{p}. 

We should not be surprised that a particular class of multifermion couplings or
corresponding Yukawa coupling models does not work. This does not means that
all possible classes of multifermion couplings or corresponding Yukawa coupling
models definitely do not work. In the ref.\cite{xue95}, we studied a possible
lattice chiral theory with an extremely large multifermion coupling ($g_1=0,
1\ll g_2<\infty$):
\begin{eqnarray}
S&\!=\!&{1\over 2a}\sum_x\left(\bar\psi^i_L(x)\gamma_\mu D^\mu\psi^i_L
(x)+\bar\psi_R(x)\gamma_\mu\partial^\mu\psi_R(x)\right)\label{action}
\\
&\!+\!&
g_2\bar\psi^i_L(x)\!\cdot\!\Delta\psi_R(x)
\Delta\bar\psi_R(x)\!\cdot\!\psi_L^i(x).\nonumber
\end{eqnarray}
This action possesses the $\psi_R$ shift-symmetry\cite{gp}.
The important consequences of the Ward identity of this symmetry are 
that\cite{xue95},
for any finite values of the multifermion coupling $g_2$,
the normal modes ($p\rightarrow 0$) of $\psi_L^i(x)$ and $\psi_R$ do not
undergo NJL spontaneous symmetry breaking for vanishing self-energy functions,
\begin{equation}
p_\mu=0,\hskip1cm \Sigma^i(p)=0, 
\label{zero} 
\end{equation} 
and do not bind to three-fermion states in the continuum limit.
It is due to these properties that we have the possibility of finding a 
segment\cite{xue95}
\begin{equation}
1\ll g_2<\infty,
\label{segment}
\end{equation}
where normal modes remain as free chiral fermions, 
only doublers ($p\simeq\pi_A$) of $\psi^i_L$ and
$\psi_R$ in (\ref{action}) are bound to form three-fermion
states: 
\begin{equation}
\Psi_R^i={1\over
2a}(\bar\psi_R\cdot\psi^i_L)\psi_R;\hskip1cm\Psi^n_L={1\over 2a}(\bar\psi_L^i
\cdot\psi_R)\psi_L^i.
\label{bound}
\end{equation}
These three-fermion states carry the appropriate quantum
numbers of the chiral gauge group $SU_L(2)$ that accommodates 
the $\psi^i_L$. These bound states (\ref{bound}) are Weyl
fermions and respectively pair up with the $\bar\psi_R$ and $\bar\psi_L^i$
to be massive, neutral $\Psi_n$ and charged $\Psi_c^i$
Dirac fermions, 
\begin{equation}
\Psi^i_c=(\psi_L^i, \Psi^i_R),\hskip1cm\Psi_n=(\Psi_L^n, \psi_R).
\label{di}
\end{equation}
And their inverse propagators are:
\begin{eqnarray}
S_c^{-1}(p)_{ij}&=&\delta_{ij}\left({i\over a}\sum_\mu\gamma_\mu \sin p^\mu P_L
+{i\over a}\sum_\mu\gamma_\mu \sin p^\mu P_R+M(p)\right),
\label{sc1}\\
S_n^{-1}(p)&=&\sum_\mu\gamma_\mu f^\mu(p)P_L
+{i\over a}\sum_\mu\gamma_\mu \sin p^\mu P_R+M(p).
\label{sn}
\end{eqnarray}
The doubler spectrum of the massive composite Dirac fermions $\Psi^i_c$ 
and $\Psi_n$
(\ref{di}) is vector-like, consistently with the $SU_L(2)\otimes U_R(1)$ chiral
symmetry. This segment (\ref{segment}) gives us a loophole to have chiral 
fermions in the continuum limit. 

In the present letter, in order to be more clear that normal modes could 
remain as
free chiral fermions, we further clarify the threshold phenomenon: the
possibility of three-fermion states turning into the virtual states of their
constituents in the low-energy limit. Provided this phenomenon occurs, we
discuss the perturbation of an $SU_L(2)$ chiral gauge interaction and gauge
anomaly. 

\vskip0.5cm
\noindent
{\bf 2.}\hskip0.3cm
We discuss how the normal modes $(p=\tilde
p\sim 0)$ of the $\psi^i_L(x)$ and $\psi_R(x)$ are possibly massless and chiral in the
low-energy limit. This is most difficult point to show for the time being,
since the strong-coupling expansion in powers of ${1\over g_2}$ breaks down
for $p\rightarrow 0$\cite{xue95}. On
the basis of the continuity \cite{nn81,ys93} of the vector-like spectrum
(\ref{sc1},\ref{sn}) of doublers in momentum space due to the locality of the theory
(\ref{action}), one may argue that the vector-like spectrum 
(\ref{sn},\ref{sc1}), which is
obtained for $p\not=0$, can be continuously extrapolated to $p\rightarrow 0$,
and we fail to have chiral fermions in the low-energy region. 

Whether the spectrum of normal modes ($p\rightarrow 0$) of $\psi^i_L$ and
$\psi_R$ is chiral, is crucially related to, whether the normal modes have been
composed to form the three-fermion states  (\ref{di}) in the segment $1\ll
g_2<\infty$. As $p\rightarrow 0$, the effective multifermion coupling for these
normal modes becomes small, as does the binding energy of these three-fermion 
states. The continuity of the spectrum (\ref{sn},\ref{sc1}) in
momentum space breaks down at the threshold where the binding energy of these
three-fermion states goes to zero. In the following discussion, we adopt the
1+1 dimensional case to illustrate this threshold phenomenon\cite{xue96}. 
 
We take the charged Dirac fermion (\ref{sc1}) on its mass shell and consider
that the time direction is continuous and one space direction is discrete. We obtain the
dispersion relation corresponding to this Dirac fermion (\ref{sc1}) for
$p\not=0$, 
\begin{equation}
E(p)=\pm\sqrt{\sin^2p+(8a^2gw^2(p))^2},
\label{des}
\end{equation}
where $E(p)$ is the dimensionless energy of the state ``$p$''. In
eq.(\ref{des}), the ``+'' sign corresponds to the dispersion relation of the
right-handed three-fermion state $\Psi_R^i(x)$. Due to the locality of the
theory, the spectrum of this bound state $\Psi^i_R(x)$ is continuous in
momentum space\cite{nn81,ys93}. The vector-like spectrum (\ref{des}) that we
obtained by the strong coupling expansion at $p\sim$ O(1), can be analytically
continued to low-momentum states $p\rightarrow 0$, unless this bound state
$\Psi^i_R$ hits the energy threshold for dissolving into its constituents.

For a given total momentum $p$ in the low-energy region, we consider a fermion
system that contains the same constituents of the bound state
$\Psi_R^i(x)$. This fermion system is the virtual state of three free chiral fermions: right-handed fermions
$\bar\psi_R$ and $\psi_R$ with momenta $p_1$ and $p_2$, and a left-handed fermion
$\psi_L^i$ with momentum $p_3$, where
\begin{equation}
p=p_1+p_2+p_3>0,\hskip0.3cm |p_i|\ll {\pi\over2},
\hskip0.3cm i=1,2,3.\label{totalm}
\end{equation}
Since the NJL spontaneous symmetry breaking does not occur 
(\ref{zero}) for the states $|p_i|\rightarrow 0$ in the segment
$(1\ll g_2<\infty)$, the total energy (continuous spectrum) of such a system is given by
\begin{eqnarray}
E_t&=&E_1(p_1)+E_2(p_2)+E_3(p_3),\nonumber\\
E_1(p_1)&=& \sqrt{\sin^2 p_1},\hskip0.5cm p_1>0,\nonumber\\ 
E_2(p_2)&=& \sqrt{\sin^2 p_2},\hskip0.5cm p_2>0,\nonumber\\ 
E_3(p_3)&=&-\sqrt{\sin^2 p_3},\hskip0.5cm p_3<0, 
\label{totale}
\end{eqnarray}
where all negative-energy states have been filled. There is no definite
relationship between the total energy $E_t$ and the total momentum $p$, since
this system is not a particle (a bound state). The total energy $E_t$ of such a 
virtual state is continuous because of the relative degrees of freedom $(p_1,p_2,p_3)$ within the
system. 

Given the total momentum $p$, the lowest energy $\min E_t$ (the threshold) and
corresponding configuration ($p_1,p_2,p_3$) 
of such a virtual state can be determined by minimizing the following total
energy with a Legendre multiplier $\lambda$ (constraint (\ref{totalm})), 
\begin{equation}
E_t=E_1(p_1)+E_2(p_2)+E_3(p_3)+\lambda p.
\label{legendre}
\end{equation}
We obtain
\begin{eqnarray}
\min E_t(p)&=&3|\sin p|,\label{thre}\\
p_3&=&-p,\hskip1cm p_1=p_2=p.\nonumber 
\end{eqnarray}

On the other hand, the three-fermion state (\ref{di}), as a
bound state, has a definite relationship between its energy and
momentum, which is given by the dispersion relation (\ref{des}) with the ``+''
sign. Given the same momentum ``$p$'' as eq.(\ref{totalm}), this bound state
is stable, if and only if there is an energy gap $\delta(p)$
(binding energy) between the threshold (\ref{thre}) and the energy (\ref{des})
of the three-fermion state, i.e., 
\begin{equation}
\delta(p)=min E_t(p)-E(p)>0.
\label{sta}
\end{equation}
The three-fermion state dissolves into its constituents, when the energy gap 
$\delta$ goes to zero,
\begin{equation}
\delta(p)=min E_t(p)-E(p)=0.
\label{con}
\end{equation}
The same discussion can be applied to the neutral three-fermion state
$\Psi_L^n$ (\ref{sn}). This discussion is very much like the case of the 
hydrogen atom,
a bound state composed of an electron and a proton, where the energy gap
between the first energy-level (n=1) and the continuous spectrum of its
virtual state is 13.6 eV.
The hydrogen atom turns into the virtual state of a free electron and a 
free proton as the energy-gap disappears ($n\gg 1$). 

Substituting eqs.(\ref{des}) and (\ref{thre}) into eq.(\ref{con}), we obtain
in the continuum limit $p\rightarrow 0$, the energy-gap closes,
\begin{equation}
\delta(p)\rightarrow 0,\hskip0.5cm p\rightarrow 0
\label{fin}
\end{equation}
where the three-fermion-state dissolves into the virtual state of three
free chiral fermions. Obviously, this plausible speculation needs to receive either a
rigorously analytical proof or a numerical evidence, which are the subject of
future work. Nevertheless, We assume there exist a threshold $\epsilon$
in momentum space.
The low-energy fermion states ``$p$'' below this threshold $\epsilon$ 
\begin{equation}
|p|<\epsilon\ll {\pi\over2},
\label{pthre}
\end{equation}
are massless and chiral. This threshold $\epsilon$ certainly depends on the
multifermion coupling $g_2$.

To end this discussion, we would like to point out that the fact that the normal
modes do not undergo NJL spontaneous symmetry breaking (\ref{zero}) for any
finite value of the multifermion coupling $g_2$ is extremely crucial. This
means, with respect to normal modes, there is no broken phase separating the
strong symmetric phase from the weak symmetric phase. In other words, there
is no a mass-gap in eq.(\ref{totale}). Otherwise, the system (\ref{totale})
would be massive, the energy gap (\ref{sta}) would never be zero and we would end up
with a vector-like spectrum in the low-energy region ($p\rightarrow 0$). This is
the main reason for the failure of EP's model, as pointed out in
ref.\cite{gpr}. Thus, there might be a chance that (i) the
chiral continuum limit can be defined at a transition from one symmetric
phase to another symmetric phase; (ii) there is a region (\ref{segment})
in the coupling space
$g_2$ where doublers are gauge-invariantly decoupled and normal modes are
chiral (non NJL-generated masses). The latter is a possible case in the segment
($1\ll g_2<\infty$), which we have discovered. 

\vskip0.5cm
\noindent
{\bf 3.}
\hskip0.3cm
Can the scaling region $1\ll g_2<\infty$
be altered, as the $SU_L(2)$ chiral gauge coupling $g$ is 
turned on in the action (1)? We expect a
slight change of the scaling segment. We should be able to re-tune the multifermion
coupling $g_2$ to compensate these changes, due the fact that the
$SU_L(2)$-chiral gauge interaction does not spoil the $\psi_R$ shift-symmetry.
As a consequence, the Ward identity associated with the
$\psi_R$ shift-symmetry remains valid when the chiral gauge interaction is
turned on.

Based on the Ward identity of the $\psi_R$ shift-symmetry (15) in 
ref.\cite{xue95}, we take
functional derivatives with respect to the gauge field $A'_\mu$, and we arrive
at the following Ward identities, 
\begin{equation}
{\delta^{(2)} \Gamma\over\delta
A'_\mu\delta\bar\psi'_R}={\delta^{(3)} \Gamma\over\delta
A'_\mu\delta\psi'_R\delta\bar\psi'_R}={\delta^{(3)} \Gamma\over\delta
A'_\mu\delta\Psi'^n_L\delta\bar\psi'_R}=\cdot\cdot\cdot=0.
\label{wa}
\end{equation}
These Ward identities and the identical vanishing of 1PI functions
containing external gauge fields $A_\mu(x)$, ``spectator'' fermion 
$\psi_R(x)$ and neutral
composite field $\Psi^n_L(x)$ show non interaction
between the gauge field and ``spectator'' fermion $\psi_R$ and 
neutral three-fermion states $\Psi_L^n(x)$. Thus, we disregard these
neutral modes.

In order to find the interacting vertex between the gauge 
boson and the charged Dirac fermion $\Psi^i_c(x)$, we need to consider the 
following three-point functions,
\begin{eqnarray}
\langle\Psi_c(x_1)\bar\Psi_c(x) A_\nu^a(y)\rangle
&=&\langle\psi_L(x_1)\bar\psi_L(x) A_\nu^a(y)\rangle
+\langle\psi_L(x_1)\bar\Psi_R(x) A_\nu^a(y)\rangle
\nonumber\\
&+&\langle\Psi_R(x_1)\bar\psi_L(x) A_\nu^a(y)\rangle
+\langle\Psi_R(x_1)\bar\Psi_R(x) A_\nu^a(y)\rangle,
\label{3points}
\end{eqnarray}
where we omit henceforth the $SU_L(2)$ indices $i$ and $j$.
Assuming the vertex functions to be $\Lambda_\mu^a(p,p')$ and $q=p'+p$,
we can write the three-point functions in momentum space:
\begin{eqnarray}
\int_{x_1xy}e^{i(p'x\!+\!px_1\!-\!qy)}\langle\psi_L(x_1)\bar\psi_L(x) A_\nu^a(y)\rangle
\!&=&\! G^{ab}_{\nu\mu}(q)S_{LL}(p) \Lambda^b_{\mu LL}(p,p')S_{LL}(p');
\label{rpa1}\\
\int_{x_1xy}e^{i(p'x\!+\!px_1\!-\!qy)}\langle\psi_L(x_1)\bar\Psi_R(x) A_\nu^a(y)\rangle
\!&=&\! G^{ab}_{\nu\mu}(q)S_{LL}(p) \Lambda^b_{\mu LR}(p,p')S_{RR}(p');
\label{rpa2}\\
\int_{x_1xy}e^{i(p'x\!+\!px_1\!-\!qy)}\langle\Psi_R(x_1)\bar\Psi_R(x) A_\nu^a(y)\rangle
\!&=&\! G^{ab}_{\nu\mu}(q)S_{RR}(p) \Lambda^b_{\mu RR}(p,p')S_{RR}(p'),
\label{rpa3}\\
\int_{x_1xy}e^{i(p'x\!+\!px_1\!-\!qy)}\langle\Psi_c(x_1)\bar\Psi_c(x) A_\nu^a(y)\rangle
\!&=&\! G^{ab}_{\nu\mu}(q)S_c(p) \Lambda^b_{\mu c}(p,p')S_c(p'),
\label{rpad}
\end{eqnarray}
where $G^{ab}_{\nu\mu}(q)$ is the propagator of the gauge boson;
$S_{LL}(p), S_{RR}(p)$ and $S_c(p)$ are the propagators of chiral fermions 
$\psi_L(x), \Psi_R(x)$ and Dirac fermion $\Psi_c(x)$.

Using the small gauge-coupling expansion, one can directly calculate the 
three-point function:
\begin{eqnarray}
&&\langle\psi_L(x_1)\bar\psi_L(x) A^a_\mu(y)\rangle=i{g\over 2}\left(
{\tau^a\over 2}\right)\sum_z\langle\psi_L(x_1)\bar\psi_L(x)\rangle\gamma_
\rho\nonumber\\
&&\left[\langle\psi_L(z\!+\!\rho)\bar\psi_L(x)\rangle\langle
A^b_\rho(z\!+\!{\rho\over2})A^a_\mu(y)\rangle\!+\!
\langle\psi_L(z\!-\!\rho)\bar\psi_L(x)\rangle\langle
A^b_\rho(z\!-\!{\rho\over2})A^a_\mu(y)\rangle\right],
\label{3p}
\end{eqnarray}
and obtain
\begin{eqnarray}
\Lambda_{\mu LL}^{(1)}(p,p') &=& ig\left(\tau^a\over 2\right)\cos\left({p+p'\over 2}
\right)_\mu\gamma_\mu P_L,\label{normal}\\
\Lambda_{\mu\nu LL}^{(2)}(p,p') &=& -i{g^2\over 2}\left(\tau^a\tau^b\over 4\right)
\sin\left({p+p'\over 2}\right)_\mu
\gamma_\mu \delta_{\mu\nu}P_L,\nonumber\\
&\cdot\cdot\cdot.&\nonumber
\end{eqnarray}

By the strong coupling expansion in powers of ${1\over g_2}$, we try to compute
the other three-point functions in eqs.(\ref{3points}) in terms of
$\langle\psi_L(x_1)\bar\psi_L(x) A_\nu^a(y)\rangle$, and we obtain the 
following recursion relations
at the leading nontrivial order, 
\begin{eqnarray}
\langle\psi_L(x_1)\bar\psi_L(x) A_\nu^a(y)\rangle
&=&{1\over g_2\Delta^2(x)}\left({1\over 2a}\right)^2\sum^\dagger_\rho
\langle\psi_L(x_1)\bar\Psi_R(x+\rho) A_\nu^a(y)\rangle\gamma_\rho
\label{ra1}\\
\langle\psi_L(x_1)\bar\psi_L(x) A_\nu^a(y)\rangle
&=&{1\over g_2\Delta^2(x_1)}\left({1\over 2a}\right)^2\sum^\dagger_\rho
\gamma_\rho\langle\Psi_R(x_1+\rho)\bar\psi_L(x) A_\nu^a(y)\rangle
\label{ra1'}\\
\langle\Psi_R(x_1)\bar\Psi_R(x) A_\nu^a(y)\rangle
&=&{1\over g_2\Delta^2(x)}\left({1\over 2a}\right)^2\sum^\dagger_\rho\gamma_\rho
\langle\psi_L(x_1)\bar\Psi_R(x+\rho) A_\nu^a(y)\rangle.
\label{ra3}
\end{eqnarray}
We make the Fourier transform in both sides of the above recursion relations
for ($p,p'\not=0$), and obtain, 
\begin{eqnarray}
S_{LL}(p) \Lambda^a_{\mu LL}(p,p')S_{LL}(p')
&=&{i\over aM(p')}
S_{LL}(p) \Lambda^a_{\mu LR}(p,p')S_{RR}(p')
\sum_\rho\sin p'_\rho\gamma^\rho
\label{ra1p}\\
S_{LL}(p) \Lambda^a_{\mu LL}(p,p')S_{LL}(p')
&=&{i\over aM(p)}\sum_\rho\sin p_\rho\gamma^\rho
S_{RR}(p) \Lambda^a_{\mu RL}(p,p')S_{LL}(p')
\label{ra1p'}\\
S_{RR}(p) \Lambda^a_{\mu RR}(p,p')S_{RR}(p')
&=&{i\over aM(p')}
\sum_\rho\sin p'_\rho\gamma^\rho S_{LL}(p)\Lambda^a_{\mu LR}(p,p')S_{RR}(p').
\label{ra3p}
\end{eqnarray}
In these equations, the propagator of the gauge boson, $G^{ab}_{\nu\mu}(q)$, is
eliminated from both sides of equations. 

Using these recursion relations (\ref{ra1p}-\ref{ra3p}) and $S_{LL}(p)$,
$S_{RR}(p)$ obtained in ref.\cite{xue95}, we can compute the
vertex functions $\Lambda^a_{\mu RL}(p,p')$, $\Lambda^a_{\mu LR}(p,p')$ and
$\Lambda^a_{\mu RR}(p,p')$ in terms of the vertex function
$\Lambda^a_{\mu LL}(p,p')$ (\ref{normal}),
\begin{eqnarray}
M(p')\Lambda^a_{\mu LL}(p,p')
&=&\Lambda^a_{\mu LR}(p,p')\left({i\over a}\right)
\sum_\rho\sin p'_\rho\gamma^\rho,
\label{v1p}\\
M(p)\Lambda^a_{\mu LL}(p,p')
&=&\left({i\over a}\right)\sum_\rho\sin p_\rho\gamma^\rho
\Lambda^a_{\mu RL}(p,p'),
\label{v1p'}\\
M(p')\Lambda^a_{\mu RR}(p,p')
&=&\left({i\over a}\right)
\sum_\rho\sin p'_\rho\gamma^\rho \Lambda^a_{\mu LR}(p,p').
\label{v3p}
\end{eqnarray}
Taking $\Lambda^a_{\mu LL}(p,p')$ to be eq.(\ref{normal}) at leading 
order, we obtain
\begin{eqnarray}
\Lambda_{\mu RR}^{(1)}(p,p') &=& ig\left(\tau^a\over 2\right)\cos\left({p+p'\over 2}
\right)_\mu\gamma_\mu P_R,\label{vr}\\
\Lambda_{\mu LR}^{(1)}(p,p')\left({i\over a}\right)
\sin p'_\mu &=& {1\over2}M(p')ig\left(\tau^a\over 2\right)
\cos\left({p+p'\over 2}
\right)_\mu ,\label{vlr}\\
\left({i\over a}\right)
\sin p_\mu\Lambda_{\mu RL}^{(1)}(p,p') &=& {1\over2}M(p)ig\left(\tau^a\over 2\right)
\cos\left({p+p'\over 2}
\right)_\mu.
\label{vrl}
\end{eqnarray}
Thus, the coupling (\ref{3points}) between the gauge field and 
Dirac fermion (\ref{sc1})
is given by
\begin{equation}
\Lambda_{\mu c}^{(1)}=\Lambda_{\mu LL}^{(1)}+\Lambda_{\mu LR}^{(1)}
+\Lambda_{\mu RL}^{(1)}+\Lambda_{\mu RR}^{(1)}.
\label{vdirac}
\end{equation}
These calculations can be straightforwardly generalized to higher orders of
the perturbative expansion in powers of the gauge coupling.
One can check that these results precisely obey the following Ward identity of
the exact $SU_L(2)$ chiral gauge symmetry ($p', p\not= 0$),
\begin{equation}
\left({i\over a}\right)(
\sin p_\mu-\sin p'_\mu)\Lambda_{\mu c}^{(1)}(p,p')=S_c^{-1}(p)-S_c^{-1}(p'),
\label{gward}
\end{equation}
where the gauge coupling $g$ and generator ${\tau_a\over 2}$ are eliminated
from the vertex $\Lambda_{\mu c}$. These results are what we expected, since we are in the symmetric phase
($1\ll g_2<\infty$) where the exact $SU_L(2)$ chiral gauge symmetry is
realized by the vector-like spectrum excluding the low-energy states $p'\not=0$
and $p\not=0$. 

When the momenta $p', p$ (\ref{gward}) reach the threshold $\epsilon$ 
(\ref{pthre}), the right-handed three-fermion state $\Psi_R^i(x)$
is supposed to dissolve into its constituents. The 1PI vertex functions $\Lambda_{\mu RR}$,
$\Lambda_{\mu RL}$, and $\Lambda_{\mu LR}$ relating to $\Psi_R^i(x)$ vanish at
this threshold. The coupling vertex (\ref{vdirac}) between the gauge boson and
fermion should turn out to be chiral consistently with the $SU_L(2)$ chiral gauge
symmetry, ($p', p\rightarrow 0$) 
\begin{equation}
\left({i\over a}\right)(
\sin p_\mu-\sin p'_\mu)\Lambda_{\mu LL}^{(1)}(p,p')
=S_{L}^{-1}(p)-S_{L}^{-1}(p'),
\label{glward}
\end{equation}
where $S_{L}^{-1}(p)$ is the propagator of the left-handed chiral fermion 
and the Ward identity is realized by the chiral spectrum. Here
we stress again that the disappearance of the three-fermion (right-handed)
state at the threshold $\epsilon$ is essential point to obtain the continuum chiral gauge coupling 
(\ref{normal}) in the
low-energy limit. However, we have to confess that similar to the threshold
(\ref{pthre}), eq.(\ref{glward}) is a plausible speculation for the time being,
since we need more evidences and computations to show whether or not it is true.

\vskip0.5cm
\noindent
{\bf 4.}
\hskip0.3cm
Given the scenario of the chiral gauge coupling and spectrum (vector-like
for $p\simeq\pi_A$ and chiral for $p\simeq 0$), 
one should expect that the gauge field should not only chirally couple
to the massless chiral fermion of the $\psi_L^i$ in the low-energy regime, but
also vectorially couple to the massive doublers of Dirac fermion $\Psi_c^i$ in
the high-energy regime. We discuss the gauge anomaly and the
renormalization of perturbative gauge theories. 

We consider the following $n$-point 1PI functional:
\begin{equation}
\Gamma^{(n)}_{\{\mu\}}=
{\delta^{(n)}\Gamma(A')\over\delta A'_{\mu_1}(x_1)\cdot\cdot\cdot\delta 
A'_{\mu_j}
(x_j)\cdot\cdot\cdot\delta A'_{\mu_n}(x_n)},
\label{fun}
\end{equation}
where $j=1\cdot\cdot\cdot n, (n\geq 2)$ and $\Gamma(A')$ is the vacuum
functional. The perturbative computation of the 1PI vertex functions
$\Gamma^{(n)}_{\{\mu\}}$ can be straightforwardly performed by adopting the
method presented in ref.\cite{smit82} for lattice QCD. Dividing the integration
of internal momenta into 16 hypercubes where the 16 modes live, we have 16
contributions to the truncated vertex functions. The region where the chiral
fermion modes of the continuum limit exist, is defined as
\begin{equation}
\Omega=[0,\epsilon]^4,\hskip0.5cm p<\epsilon\ll {\pi\over 2},\hskip0.5cm
p\rightarrow 0, 
\label{continuum}
\end{equation}
where $\epsilon$ is the energy-threshold given by (\ref{pthre}),
on which $\Psi_R(x)$ disappears.

As a first example, we deal with the vacuum polarization
\begin{equation}
\Pi_{\mu\nu}(p)=\sum_{i=1}^{16}\Pi^i_{\mu\nu}(p),\hskip0.5cm
\Pi^d_{\mu\nu}(p)=\sum_{i=2}^{16}\Pi^i_{\mu\nu}(p).
\label{vacuum}
\end{equation}
For the contributions
$\Pi^d_{\mu\nu}(p)$ from the 15 doublers $(i=2,...,16)$, we make a Taylor
expansion in terms of external physical momenta $p=\tilde p$ and the following
equation is, {\it mutatis mutandis}, valid 
\cite{smit82},
\begin{eqnarray}
\Pi^d_{\mu\nu}(p)&=&\Pi^\circ_{\mu\nu}(0)+\Pi^{d(2)}_{\mu\nu}(p)(\delta_{\mu\nu}
p^2-p_\mu p_\nu)\nonumber\\
&+&\sum^{16}_{i=2}\left(1-p_\rho|_\circ\partial_\rho-{1\over2}
 p_\rho p_\sigma|_\circ\partial_\rho\partial_\sigma\right)
\Pi^{con}_{\mu\nu}(p,m_i),
\label{smit}
\end{eqnarray}
where $|_\circ f(p)=f(0)$ and $m^i$ are doubler masses. The first and second terms are specific for the
lattice regularization. Since the 15 doublers are gauged as an $SU(2)$ QCD-like
gauge theory with propagator (\ref{sc1}) and interacting vertex (\ref{vdirac}), the
Ward identities (\ref{gward}) associated with this vectorial gauge symmetry result in the
vanishing of the first divergent term $\Pi^\circ_{\mu\nu}(0)$ and the gauge
invariance of the second finite contact term in eq.(\ref{smit}). We recall
that in the Rome approach, this was achieved by enforcing Ward identities and 
gauge-variant counterterms. The third term
in eq.(\ref{smit}) corresponds to the relativistic contribution of the 15
doublers. The $\Pi^{con}_{\mu\nu}(p,m_i)$ is logarithmically divergent and
evaluated in some continuum regularization. For doubler masses $m_i$ of
$O(a^{-1})$, the third term in eq.(\ref{smit}) is just finite and
gauge-invariant. 

We turn to the contribution $\Pi^n_{\mu\nu}(p)$ of the normal
mode that is in the first hypercube $\Omega=[-\epsilon,\epsilon]^4$
(\ref{continuum}).
Based on the plausible speculation that the normal mode and gauge
coupling are chiral, we can use some regularization to calculate this 
contribution, 
\begin{equation}
\Pi^n_{\mu\nu}(p)=\Pi^{n(2)}_{\mu\nu}(p)(\delta_{\mu\nu}
p^2-p_\mu p_\nu).
\label{nsmit}
\end{equation}
The Ward identity (\ref{glward})
associated with this chiral gauge symmetry render eq.(\ref{nsmit}) gauge
invariant. The $\epsilon$-dependence ($\ln\epsilon$) in eq.(\ref{nsmit}) has to be exactly
cancelled by those contributions (\ref{smit}) from doublers, because of
the continuity of 1PI vertex functions in momentum 
space. 
In summary, the total vacuum polarization $\Pi_{\mu\nu}(p)$ contains two parts:
(i) the vacuum polarization of the normal mode (chiral) of $\psi^i_L$ in some continuum 
regularization; (ii) gauge invariant finite terms stemming from doubler
contributions. The second part is the same as perturbative lattice 
QCD, and can be subtracted away in a normal renormalization prescription.

The second example is the 1PI vertex functional $\Gamma^{(n)}_{\{\mu\}}(\{p\}) (n\geq
4)$,
\begin{eqnarray}
\Gamma^{(n)}_{\{\mu\}}(\{p\})&=&\sum^{16}_{i=1}
\Gamma^{(n)i}_{\{\mu\}}(\{p\},m_i)\hskip0.5cm n\geq 4,
\label{n=4}\\
\{p\}&=&p_1,p_2,\cdot\cdot\cdot\nonumber\\
\{\mu\}&=&\mu_1,\mu_2,\cdot\cdot\cdot,\nonumber
\end{eqnarray}
where the internal momentum integral is analogously divided into the contributions
from sixteen sub-regions of the Brillouin zone where the sixteen modes live. Based on gauge invariance
and power counting, one concludes that up to some gauge invariant finite terms,
the $\Gamma^{(n)}_{\{\mu\}}(\{p\}) (n\geq 4)$ (\ref{n=4}) contain the 15
continuum expressions for 15 massive ($m_i$) Dirac doublers and one for the
massless chiral mode. The 15 doubler contributions vanish for $m_i\sim
O(a^{-1})$. The $n$-point 1PI vertex functions (\ref{n=4}) end up with their
continuum counterpart for the chiral fermion and some gauge invariant finite
terms. These gauge invariant finite terms, which come from  doubler 
contributions,
are similar to those in Wilson lattice QCD, and can be subtracted away in the
normal renormalization prescription. 

The most important contribution to the vacuum functional is the triangle graph
$\Gamma_{\mu\nu\alpha} (p,q)$, which is linearly divergent. Again, dividing the
integration of internal momenta into 16 hypercubes, one obtains\cite{smit82}
\begin{eqnarray}
\Gamma_{\mu\nu\alpha}(p,q)&=&\sum^{16}_{i=1}\Gamma^i_{\mu\nu\alpha}(p,q)
\nonumber\\
\Gamma^i_{\mu\nu\alpha}(p,q)&=&\Gamma^{i(\circ)}_{\mu\nu\alpha}(0)+
p_\rho\Gamma^{i(1)}_{\mu\nu\alpha ,\rho}(0)
+q_\rho\Gamma^{i(1)}_{\mu\nu\alpha ,\rho}(0)\nonumber\\
&+&\left(1-|_\circ - p_\rho|_\circ\partial_\rho - q_\rho|_\circ\partial_\rho
\right)
\Gamma^{con}_{\mu\nu\alpha}(p,q,m_i),
\label{smit1}
\end{eqnarray}
where $\Gamma^{con}_{\mu\nu\alpha}(p,q,m_i)$ is evaluated in some continuum
regularization. As for the 15 contributions of Dirac doublers
($i=2\cdot\cdot\cdot 15$), the first three terms in eq.(\ref{smit1}) 
are zero owing to the vector-like
Ward identity (\ref{gward}). The non-vanishing contributions are similar to
those in Wilson lattice QCD.
These contributions are gauge-invariant and finite (as $m_i\sim O(a^{-1})$),
thus, disassociate from the gauge anomaly. 

The non-trivial contribution of the
chiral mode in the hypercube $\Omega =[-\epsilon,\epsilon]^4$ is given by
\begin{eqnarray}
\Gamma^{i=1}_{\mu\nu\alpha}(p,q)\!&\!=\!&\!\int_\Omega\!
{d^4k\over(2\pi)^4}\tr\!\left[S(k\!+\!{p\over2})\Gamma_\mu(k)S(k\!-\!{p\over2})
\Gamma_\nu(k\!-\!{p\!+\!q\over2})S(k\!-\!{p\over2}\!-\!q)\Gamma_\alpha(k\!-\!
{q\over2})\right]\nonumber\\
&&+(\nu\leftrightarrow\alpha),
\label{tri}
\end{eqnarray}
where $S(p)=S_L(p)$ is the propagator of the left-handed chiral fermion, and vertex $\Gamma_\mu$ 
is given by eq.(\ref{normal}). Other contributions containing anomalous vertices
$(\psi\bar\psi AA, \psi\bar\psi AAA)$ vanish within 
the hypercube $\Omega =[-\epsilon,\epsilon]^4$. As is well known, eq.~(\ref{tri}) 
is not gauge invariant. To evaluate eq.(\ref{tri}),
one can use some continuum regularization. As a result, modulo possible local
counterterms, we obtain the consistent gauge anomaly for the non-abelian 
chiral gauge theories as the continuum one: 
\begin{equation}
\delta_g\Gamma(A')=-{ig^2\over24\pi^2}\int d^4x
\epsilon^{\alpha\beta\mu\nu}\tr\theta_a(x)\tau_a\partial_\nu 
\left[A_\alpha(x)\left(\partial_\beta A_\mu+{ig\over2}A_\beta (x)A_\mu(x)
\right)\right],
\label{anomaly}
\end{equation}
where the gauge field $A_\mu={\tau^a\over2} A^a_\mu$. 
The $SU_L(2)$ chiral gauge theory is anomaly-free for
$\tr(\tau^a,\{\tau^b,\tau^c\})=0$, and the gauge current
\begin{equation}
J^a_\mu=i\bar\psi_L\gamma^\mu{\tau^a\over2}\psi_L={\delta\Gamma(A)\over \delta
A_\mu^a(x)},\hskip0.5cm \partial^\mu J^a_\mu=0,
\label{conser}
\end{equation}
is covariantly conserved and gauge invariant. It is surprising that we can
consistently obtain the correct chiral gauge anomaly (\ref{anomaly}) in a
speculative scenario, where the normal mode is chiral and doublers are massive and
vector-like. What is interesting is that in this gauge invariant scenario and
action (\ref{action}), we consistently achieve the chiral gauge anomaly, since
this is different from the general idea that the origin of chiral gauge
anomalies is due to the explicit breaking of chiral gauge symmetries at the
cutoff. In the following paragraph, we give some discussion on this point. 

A most subtle property of the naive lattice chiral gauge theory is the
appearance of 16 modes. Each mode produces the chiral gauge anomaly with
definite axial charge $Q_5$\cite{smit82}, such that the finite (regularized)
theory is anomaly-free and the chiral gauge symmetry is perfectly preserved. As
has been seen, the 15 doublers are decoupled as massive Dirac fermions that are
vector-gauge-symmetric (\ref{gward}). Thus, they decouple from the chiral gauge
anomaly as well. Only the anomaly associated with the normal (chiral) mode of
the $\psi^i_L$ is left and is the same as the continuum one. The condition for
this circumstance to occur is the disappearance of the right-handed
three-fermion state $\Psi^i_R$ in the low-energy limit. Otherwise, the chiral
gauge anomaly (\ref{anomaly}) from the normal mode of $\psi_L^i$ would be exactly cancelled by
that from $\Psi^i_R$. In future work, we will proceed to more detailed 
discussion on this
point by looking at how this three-fermion state flows, dissolves into its
constituents and fills up the Dirac sea\cite{nc}.  

It seems surprising that we start from a gauge symmetric action and end up
with the correct gauge anomaly. Normally, one may claim that the anomaly has to
come from the explicit breaking of the chiral gauge symmetry in a regularized
action (e.g., the Wilson term). This statement is indeed correct if regularized
actions are bilinear\footnote{ This means no multifermion couplings (Yukawa
couplings) and only gauge couplings.} in fermion fields, since this is nothing
but what the ``no-go" theorem asserts. However, we run into the dilemma that
the chiral gauge anomaly is independent of any explicit breaking parameters
(e.g., the Wilson parameter $r$ and fermion masses). In fact, the most
essential and intrinsic {\it raison d'\^etre} of producing the correct chiral
gauge anomaly is ``decoupling doublers'' rather than ``explicit breaking of
chiral gauge symmetries''. If we adopt a bilinear action (e.g.,~the Wilson
action) to decouple doublers, we must explicitly break chiral gauge symmetry,
which is just a superficial artifact in bilinear actions. However, if we give
up the bilinearity of regularized actions in fermion fields and find a
chiral-gauge-invariant way (as the scenario we discussed) to decouple doublers,
we should not be surprised to achieve the correct gauge anomaly (\ref{anomaly}). 

The Ward identities (\ref{gward}) and (\ref{glward}) play an extremely
important role to guarantee that the gauge perturbation theory in the scaling
region ($1\ll g_2<\infty$) is gauge symmetric. To all orders of the gauge coupling
perturbation theory, gauge boson masses vanish and the gauge boson propagator
is gauge-invariantly transverse. The gauge perturbation theory can be described
in the normal renormalization prescription as that of the QCD and QED theory.
In fact, due to the manifest $SU_L(2)$ chiral gauge symmetry and corresponding
Ward identities that are respected by the spectrum (vector-spectrum for
$p\not=0$ and chiral-spectrum for $p=0$) in this possible scaling regime, we
should then apply the Rome approach \cite{rome} (which is based on the
conventional wisdom of quantum field theories) to perturbation theory in the
small gauge coupling. It is expected that the Rome approach should work in the
same way but all gauge-variant counterterms are prohibited. 

However, we cannot even speculate anything about the aspect of non-perturbative
dynamics of the $SU_L(2)$ chiral gauge theory in this scaling segment, since
the whole computation is strictly perturbative in the gauge coupling. The
anomalous global symmetries are not involved in this paper, and readers are
referred to the ref.\cite{cx}. 

I thank Profs.~G.~Preparata, M.~Creutz, H.B.~Nielsen and E.~Eichten for
discussions, and P.~Ratcliffe for reading the manuscript.

\end{document}